\documentstyle[12pt,german]{article}

\input amssym.def
\input amssym

\begin{document}

\author{R. Beutler\thanks{%
Permanent Address : S\"{u}ddeutscher Rundfunk, Rundfunk-Versorgungsplanung,
Neckar\-strasse 230, 70190 Stuttgart, Germany} and B.G. Konopelchenko\thanks{%
Also : Budker Institute of Nuclear Physics, Novosibirsk - 90 , Russia} \\
Universit\`{a} degli Studi di Lecce,\\
Dipartimento di Fisica, \\
73100 Lecce, Italy}
\title{Surfaces of Revolution via the Schr\"{o}dinger Equation : Construction,
Integrable Dynamics and Visualization }
\date{}
\maketitle

\begin{abstract}
Surfaces of revolution in three-dimensional Euclidean space are considered.
Several new examples of surfaces of revolution associated with well-known
solvable cases of the Schr\"{o}dinger equation (infinite well, harmonic
oscillator, Coulomb potential, Bargmann potential, etc.) are analyzed and
visualized. The properties of such surfaces are discussed. Two types of
deformations (evolutions) of the surfaces of revolution, namely 1)
preserving the Gaussian curvature and 2) via the dynamics of the Korteweg-de
Vries equation are discussed.
\end{abstract}

\section{Introduction}

Surfaces and curvature are the key ingredients in a number of problems and
phenomena in mathematics and physics. The differential geometry of surfaces
in three-dimensional Euclidean space $R^3$ has been completed, in essence,
at the end of last and the beginning of this century ( see e.g. \cite{1}-%
\cite{3} ). Basic differential equations which describe surfaces in $R^3$
have been studied from various points of view. A study of the interrelation
between special classes of surfaces and linear and nonlinear differential
equations ( PDE ) has been one of the classical problems of differential
geometry \cite{1}-\cite{3}.

Interfaces, surfaces, fronts and their dynamics are also the key objects in
numerous nonlinear phenomena in classical physics, such as surface waves,
growth of crystals, propagation of flame fronts, deformations of membranes
and many problems of hydrodynamics ( see e.g. \cite{4}, \cite{5} ). In
quantum field theory and statistical physics surfaces are of importance,
too. Numerous papers have been devoted to the study and applications of
integrals over surfaces in gauge field theories, string theory, quantum
gravity and statistical physics (see e.g. \cite{6}, \cite{7}).

The discovery of the inverse scattering ( spectral ) transform ( IST\ )
method, a powerful tool to construct and solve integrable PDEs ( see e.g. 
\cite{8}-\cite{10} ), about 30 years ago, gave a new impulse for the study
of the interrelations between nonlinear PDEs and differential geometry of
surfaces. On one hand, new nonlinear integrable PDEs ( in addition to the
sine-Gordon and Liouville equations ) with a particular geometrical meaning
have been found ( see e.g. \cite{11}-\cite{13} ). On the other hand, the IST
method has been used to construct wide classes of surfaces ( see e.g. \cite
{14}- \cite{15a} ).

An alternative use of the IST method in the framework of differential
geometry of surfaces has been proposed recently in the papers \cite{16}-\cite
{18}. The approach discussed in \cite{16} allows to generate surfaces in $%
R^3 $ ( and $R^N$ ) via two-dimensional linear problems and formulate their
integrable dynamics via the associated 2+1-dimensional integrable PDEs. This
approach has been applied in 2-dimensional gravity \cite{19} and in some
pure mathematical problems of differential geometry \cite{20}. The approach
discussed in \cite{17} and \cite{18} is quite different. It is mainly
concerned with the intrinsic geometry of surfaces and spaces. In \cite{17}
it was noted, in particular, that the recent results on exact solutions of
the one-dimensional Schr\"{o}dinger equation and the Korteweg-de Vries ( KdV
) equation provide us with the variety of surfaces where metric and Gaussian
curvature are given by simple explicit formulas. Surfaces of revolution form
a very special and simple class of surfaces. Nevertheless, they are of great
importance both in mathematics and in physics. Perhaps, they are the best
candidates to deal with in terms of the approach proposed in \cite{17} and 
\cite{18}, also since in the literature only few examples of surfaces of
revolution ( spherical and pseudospherical surfaces of constant Gaussian
curvature, catenoid ) are discussed.

In the present paper we consider surfaces of revolution in three-dimensional
Euclidean space $R^3$ and their integrable deformations ( dynamics ). Our
approach is based on the observation that surfaces of revolution in $R^3$
are completely governed by the one-dimensional Schr\"{o}dinger equation in
the sense that the Gauss equation for surfaces in terms of geodesic
coordinates, i.e. with a metric

\begin{equation}
ds^2=dx^2+H^2d\varphi ^2,  \label{e1.1a}
\end{equation}

simply reads

\begin{equation}
H_{xx}+KH=0  \label{e1.1}
\end{equation}

where $K$ is the Gaussian curvature. It is used, for instance, to analyze
surfaces of constant Gaussian curvature (see e.g. \cite{1}-\cite{3}). The
connection of (\ref{e1.1}) with the standard one-dimensional Schr\"{o}dinger
equation

\begin{equation}
-\Psi _{xx}+u\Psi =E\Psi  \label{e1.2}
\end{equation}

is established by the simple relations

\begin{equation}
K=E-u\;,\;H= Re(\Psi )\;(\;{\rm or\;}H=Im(\Psi )\;).
\label{e1.2a}
\end{equation}

In this paper we will analyze and visualize the surfaces of revolution
associated with well-known solvable cases of equation (\ref{e1.2}). We treat
the infinite well potential, $\delta $-function potential, harmonic
oscillator, effective radial Coulomb potential and Bargmann potentials. Such
''physically motivated'' surfaces of revolution have several properties
which distinguish them from the old ''mathematically motivated'' surfaces,
like the catenoid. In particular, surfaces of revolution associated with the
bound states are spindle-shaped surfaces which at infinity tend to strings.
The surfaces generated by the $n+1$-th excited bound state have $n$ knots (
conic points ).

We will consider also the deformations ( dynamics ) of surfaces of
revolution of two distinct types. The evolutions of the first type preserve
the Gaussian curvature $K(x)$ and change the metric according to

\begin{equation}
\Psi (x)\longrightarrow \Psi ^{\prime }(x,t)=A(t)\Psi (x)+B(t)\widetilde{%
\Psi }(x)  \label{e1.3}
\end{equation}

where $A$ and $B$ are two arbitrary functions of the parameter $t$ and $%
\widetilde{\Psi }(x)$ is the solution of the Schr\"{o}dinger equation (\ref
{e1.2}) linear independent to $\Psi (x)$. The simplest case $B\equiv 0$
corresponds just to the rescaling of the metric. But even this simple
transformation can deform the surface drastically. For instance, for an
increasing function $A(t)$ the evolution in $t$ may generate discontinuities
on the surface for certain $t$ and, finally, may lead to the splitting of
the surface into disconnected parts. The deformations with $B\neq 0$ create
edges and singularities in the profile of the surface in $R^3$ immediately,
i.e. for any $t\neq 0$. Regular surfaces of revolution associated with the
bound states are absolutely instable with respect to such deformations. The
particular class of deformations for which

\begin{equation}
A=\frac 1{\sqrt{-2T_t(t)}}\;,\;B=\frac{T(t)}{\sqrt{-2T_t(t)}}  \label{e1.3a}
\end{equation}

where $T(t)$ is an arbitrary function of $t$ and $T_t$ indicates the
derivative with respect to $t$, corresponds to Liouville type evolutions
recently discussed in \cite{21}. In this case the function $q=-2\ln \Psi $
obeys the equation $q_{xt}=\exp q$.

Another type of deformation is of completely different nature. It is given
by the KdV equation for the Gaussian curvature

\begin{equation}
K_t+K_{xxx}+6KK_x-6EK_x=0  \label{e1.4}
\end{equation}

and by the equation

\begin{equation}
H_t+H_{xxx}-6EH_x-3\frac{H_xH_{xx}}H=0  \label{e1.5}
\end{equation}

for the metric. The KdV equation is integrable by the IST method and has a
number of very interesting properties ( see e.g. \cite{8}-\cite{10} ).
Consequently, the KdV-type transformations of surfaces inherits all these
properties. In particular, it preserves an infinite set of integral
characteristics of the surface of the type

\begin{equation}
Q_n=2\pi \int_{-\infty }^\infty dxC_n(x)  \label{e1.5a}
\end{equation}

where the $C_n$ are differential polynomials of $K$ ( e.g. $C_1=K$, $C_2=K^2$%
, $C_3=K_x^2-2K$ ). In contrast to the deformations of the first type the
KdV-type deformations tend to smooth out the surfaces of revolution. They
described by explicit formulas for the so-called multi-soliton solutions of
the KdV equation.

The paper is organized as follows. In section 2 we present the basic
formulas for the surfaces of revolution. The following sections 3-8 contain
the discussion of the surfaces generated by well-known solvable potentials
of the Schr\"{o}dinger equation. Then the two types of deformations of
surfaces of revolution are introduced. Their properties are discussed for
the examples treated in the preceding sections.

\section{Surfaces of Revolution : Basic Formulas and Inducing via the
Schr\"{o}dinger Equation}

Here we will present for convenience some basic formulas for the surfaces of
revolution in $R^3$ (see e.g. \cite{1}- \cite{3}). A surface of revolution
is a surface which is obtained by the rotation of a plane curve around an
axis belonging to the same plane. Denoting the Cartesian coordinates in $R^3$
as $X$, $Y$ and $Z$ one can define a surface of revolution according to

\begin{equation}
X=r\cos \varphi \;,\;Y=r\sin \varphi \;,\;Z=f(r)  \label{e2.1}
\end{equation}

where $r^2=X^2+Y^2$, $\varphi $ is the angle in the $X$-$Y$ plane and $f$ is
a arbitrary function of $r$. The choice of the function $f(r)$ specifies the
surface of revolution. The Euclidean metric in $R^3$ induces on the surface (%
\ref{e2.1}) the following metric

\begin{equation}
\Omega =[1+f_r^2(r)]dr^2+r^2d\varphi ^2  \label{e2.2}
\end{equation}

where $f_r=\frac{df}{dr}$. Introducing the variables $x$ and $H$ via

\begin{equation}
dx=\sqrt{1+f_r^2(r)}dr  \label{e2.3}
\end{equation}

and

\begin{equation}
r=H\left( x\right)  \label{e2.4}
\end{equation}

one converts the metric (\ref{e2.2}) into the form

\begin{equation}
\Omega =dx^2+H^2(x)d\varphi ^2.  \label{e2.5}
\end{equation}

The lines $\varphi ={\rm const}$ are the meridians, which are geodesics, and
the lines $x={\rm const}$ are the parallels of the surface of revolution.
The meridians and parallels are also curvature lines. For surfaces of
revolution the metric depends only on $x$. The corresponding Gauss equation
reads

\begin{equation}
H_{xx}+KH=0  \label{e2.6}
\end{equation}

where $K\left( x\right) $ is the Gaussian curvature. The second fundamental
form and other characteristics also have a very simple structure in the
geodesic coordinates $x$ and $\varphi $.

Locally any form (\ref{e2.5}) corresponds to a surface of revolution in $R^3$%
. If one considers the global properties of surfaces of revolution, however,
it is necessary to pay attention to the following. Equations (\ref{e2.3})
and (\ref{e2.4}) imply that

\begin{equation}
\left( 1-H_x^2\right) dx^2=dZ^2  \label{e2.7}
\end{equation}

or

\begin{equation}
dZ=\sqrt{1-H_x^2}dx  \label{e2.8}
\end{equation}

where $H_x=\frac{dH}{dx}$. Given $H\left( x\right) $ one gets from (\ref
{e2.8})

\begin{equation}
Z=\int_{x_0}^xdx^{\prime }\sqrt{1-H_{x^{\prime }}^2(x^{\prime })}.
\label{e2.12}
\end{equation}

This formula together with (\ref{e2.4}) defines the surface of revolution in 
$R^3$ according to (\ref{e2.1}). However, the constraint

\begin{equation}
\left| H_x\right| \leq 1  \label{e2.9}
\end{equation}

for the values of the coordinate $x$ has to be taken into account. In other
words, (\ref{e2.9}) defines the set of admissable values of $x$ while (\ref
{e2.4}) determines the corresponding set of admissable values of $r$. We
will construct the surfaces in such a way that there will be no gaps in the
profile of the surface along the $Z$-axis. Thus, in the case when $x$ varies
in the set of disconnected intervals we join the surface at the end of the
intervals in order to avoid gaps.

It is clear, however, that in general depending on the choice of the
explicit form of the metric $H$ the profile of the surface of revolution
thus may exhibit discontinuities or singularities connected to the existence
of different intervals of admissable values of $x$. The weakest irregularity
consists of edges which are due to a jump in $H_x$ but still satisfying the
constraint (\ref{e2.9}). The points for which $H_x=\pm 1$ have to be
considered carefully. At these points we have

\begin{equation}
\frac{dZ}{dx}=0  \label{e2.9a}
\end{equation}

and

\begin{equation}
\frac{dZ}{dr}=\frac{\sqrt{1-H_x^2}}{H_x}=0.  \label{e2.9b}
\end{equation}

The points under consideration do not belong to the surface, but if $H$ is
finite it is possible to connect the surface by assigning appropriate
values. This can give rise to the situation that the tangent plane at these
points on the surface is parallel to the $X$-$Y$-plane. If $d^2Z/dx^2$
changes sign at these points the surface of revolution will have cuspidal
edges. Furthermore, if $|H_x|\geq 1$ for a whole interval
of $x$ the surface will be in general disconnected by a gap parallel to the $%
X$-$Y$-plane. Note, that we will use the term cuspidal edge also for the
boundary of finite surfaces having the property (\ref{e2.9a}).

Well-known examples of surfaces of revolution are the spherical ($K=1$) and
the pseudospherical ($K=-1$) surfaces and the catenoid ( $K=-a^2/(a^2+x^2)^2$%
, $H^2=a^2+x^2$). Some other examples can be found in \cite{22}.

In our approach to the surfaces of revolution we will follow the papers \cite
{17} and \cite{18}. We start with the presentation of the Gauss equation (%
\ref{e2.6}) in the form of the standard one-dimensional Schr\"{o}dinger
equation

\begin{equation}
-\Psi _{xx}+u\left( x\right) \Psi =E\Psi .  \label{e2.10}
\end{equation}

For a given solution of (\ref{e2.10}) with fixed $u\left( x\right) $, energy 
$E$ and wavefunction $\Psi \left( x\right) $ one gets a solution of the
Gauss equation (\ref{e2.6}) via the identification

\begin{equation}
K=E-u\left( x\right) \;,\;H=Re(\Psi \left( x\right) )\;(\;{\rm or\;}H=%
Im(\Psi \left( x\right) )\;).  \label{e2.11}
\end{equation}

Our selection of surfaces of revolution is motivated by the physics
associated with the Schr\"{o}dinger equation (\ref{e2.10}). We will consider
the exactly solvable cases and mainly the corresponding bound states, i.e.
stationary states of the discrete spectrum. Cases when the potential $%
u\left( x\right) $ has singularities will be discussed as well. More
complicated problems like the scattering problems and surfaces generated by
Schr\"{o}dinger equations with random potentials will be considered
elsewhere.

All figures presented here have been made using the plotting facilities of
MAPLE. Since the integration in (\ref{e2.12}) can in general not be
calculated in closed form a cubic spline approximation for the integrand has
been used to get an polynomial approximation of the function $Z(x)$ in the
interval of interest which can always be calculated to arbitrary accuracy 
\cite{24}.

\section{Spherical and Pseudospherical Surfaces via Free Motion}

We start with the simplest possible case for the Schr\"{o}dinger equation,
i.e. the free motion. For $u\equiv 0$ and positive energy, $E>0$, the
general form of the associated metric is

\begin{equation}
H=A\cos (kx+\alpha ),  \label{e3.1}
\end{equation}

where $k^2=E$ and $A$ and $\alpha $ are arbitrary real constants. Without
loss of generality one can put $\alpha =0$. In this case we have

\begin{equation}
Z=\int^xdx^{\prime }\sqrt{1-A^2k^2\sin ^2\left( kx\right) }.  \label{e3.2}
\end{equation}

Therefore three different classes of surfaces are possible.

\begin{itemize}
\item[1.]  In the case $Ak=1$ one has

\begin{equation}
Z=\frac 1k\sin \left( kx\right) +Z_0\;\;,\;\;r=H=\frac 1k\cos \left(
kx\right) ,  \label{e3.2a}
\end{equation}

leading to

\begin{equation}
X^2+Y^2+\left( Z-Z_0\right) ^2=\frac 1{k^2}  \label{e3.2b}
\end{equation}

and our surface is the standard sphere of radius $\frac 1k$. Due to the fact
that $H$ according to (\ref{e3.1}) is periodic in $x$ we will get an
infinite number of spheres glued to each other along the $Z$-axis (figure 1).

\item[2.]  For $Ak<1$ there is also no constraint on the values of $x$. The
solution of (\ref{e3.2}) is given in terms of elliptic functions. The
surface looks like an infinite sequence of pressed spheres similar to
american footballs, which are depicted in figure 2.

\item[3.]  If we choose $Ak>1$ the variable $x$ is allowed to vary within
the intervals defined by the inequality $\left| \sin (kx)\right| \leq \frac
1{Ak}$. The surface consists of an infinite sequence of pieces from which a
single surface is shown in figure 3.
\end{itemize}

We see that the form of these surfaces depends crucially on the choice of
the normalization constant $A$. In the first and the second case we have a
regular surface while in the third case the surface has cuspidal edges which
disconnect it into distinct finite pieces.

The situation is quite different for the nonphysical case of negative energy 
$E=-\lambda ^2<0$. With this choice there are three different solutions of
equation (\ref{e2.10}), namely

\begin{eqnarray}
H_1 &=&A_1\cosh \left( \lambda x\right) ,  \nonumber  \label{e3.3} \\
H_2 &=&A_2\sinh \left( \lambda x\right) ,  \label{e3.3} \\
H_3 &=&A_3\exp \left( -\lambda x\right)  \nonumber
\end{eqnarray}

where $A_1$, $A_2$ and $A_3$ are arbitrary real constants. In the first and
the second case the corresponding surfaces exhibit edges independent of the
choice of $A_1$ and $A_2$ at the points $x_1=\pm \frac 1\lambda {\rm arcsinh}%
\left( 1/A_1\lambda \right) $ and $x_2=\pm \frac 1\lambda {\rm arccosh}%
\left( 1/A_2\lambda \right) $, respectively. They are presented in figure 4
and 5. In the third case, however, the corresponding surface has only one
edge at the point $x_3=\frac 1\lambda \ln \left| A_3\lambda \right| $.
Shifting $x\rightarrow x+x_3$ one gets the surface defined for $x>0$ with
the cuspidal edge at $x=0$ and with $H_3=\frac 1\lambda \exp \left( -\lambda
x\right) $, which is shown in figure 6. It is nothing but the famous
pseudosphere. It is generated by the rotation of the so-called tractrix.

All surfaces presented in this section are well-known. They correspond to
spherical ($E>0$) and pseudospherical ($E<0$) surfaces of different types,
which hence all are generated by the physically simplest case of the
Schr\"{o}dinger equation, namely, free motion.

\section{Surfaces of Revolution via the Infinite Well Potential}

Let us consider now the simplest nontrivial potential, namely the infinite
well potential

\begin{equation}
u\left( x\right) =\left\{ 
\begin{array}{ccc}
0 & , & 0\leq x\leq \pi a \\ 
&  &  \\ 
\infty & , & x<0\;{\rm and\;}x>\pi a,
\end{array}
\right.  \label{e4.1}
\end{equation}

where $a$ is a constant. In this case one has bound states with energies

\begin{equation}
E_n=\frac{n^2}{a^2}\;\;,\;\;n=1,2,3,...  \label{e4.2}
\end{equation}

and the corresponding wavefunctions \cite{23}

\begin{equation}
\Psi _n\left( x\right) =\left\{ 
\begin{array}{ccc}
A_n\sin \left( \frac{nx}a\right) & , & 0\leq x\leq \pi a \\ 
&  &  \\ 
0 & , & x<0\;{\rm and\;}x>\pi a,
\end{array}
\right.  \label{e4.3}
\end{equation}

where $A_n$ are arbitrary real constants, since we do not require the
normalization of the wavefunction. The choice of $u\left( x\right) $
according to (\ref{e4.1}) leads to an infinite family of surfaces whose
metric is given by

\begin{equation}
H_n=\left\{ 
\begin{array}{ccc}
A_n\sin \left( \frac{nx}a\right) & , & 0\leq x\leq \pi a \\ 
&  &  \\ 
0 & , & x<0\;{\rm and\;}x>\pi a.
\end{array}
\right.  \label{e4.5}
\end{equation}

Since $K=E-u\left( x\right) $ our surfaces have constant Gaussian curvature $%
K_n=\frac{n^2}{a^2}$ for $0\leq x\leq \pi a$ and infinite negative curvature
outside this interval. Therefore within the infinite well they look like the
surfaces with positive Gaussian curvature described in the previous section
while outside they have the form of infinitely thin tubes, i.e. strings
going to infinity.

The ground state $n=1$ gives rise to three different surfaces depending on
the choice of $A_1$. For $A_1=a$ we have the surface which consists of the
standard sphere of radius $a$, connected with the infinities by strings from
the south and north poles, respectively (left upper picture in figure 7). If 
$A_1<a$ we have surfaces of the type shown in the left upper picture of
figure 8, while the case $A_1>a$ is depicted in the left upper part of
figure 9. It has to be noted, however, that in this case at the points where 
$\left| \Psi _x\right| =1$ the modulus of the wavefunction is not equal to
zero. Gluing the strings, whose radius is zero, thus introduces a jump of
the radius.

For excited states ($n>1$) the corresponding surfaces of revolution look
like a chain of $n$ pieces of the form as for $n=1$ separated by $n-1$ knots
(conic points) and connected by strings to infinity. If we choose $A_n=\frac
an$ we get a chain of spheres with radius $\frac an$ (figure 7), for $%
A_n<\frac an$ the chain consists of squeezed spheres (figure 8) which are
extended to infinity by the strings. The choice $A_n>\frac an$ is connected
to a sequence of squeezed and cut spheres separated by cuspidal edges. As
before, this is accompanied by a jump of the radius to zero for the strings
tending to infinity (figure 9).

It should be noted that if one chooses for the amplitude $A_n$ the physical
normalization of the function $\Psi _n$ ($\int_{-\infty }^\infty dx\Psi
_n^2=1$), i.e. $A_n=\sqrt{\frac 2{\pi a}}$ then the n-th excited state gives
rise to the sequence of spheres if $\pi a^3=2n^2$.

\section{$\delta $-Function Potentials and Surfaces with Singular Gaussian
Curvature}

The next of our examples is determined by the potential

\begin{equation}
u\left( x\right) =-2\varkappa \delta \left( x\right) ,  \label{e5.1}
\end{equation}

where $\varkappa $ is a positive real constant and $\delta \left( x\right) $
is the Dirac delta-function. There is one bound state with

\begin{equation}
E=-\varkappa ^2  \label{e5.2}
\end{equation}

and the corresponding wavefunction reads \cite{23}

\begin{equation}
\Psi \left( x\right) =Ae^{-\varkappa \left| x\right| }.  \label{e5.3}
\end{equation}

For $A\varkappa <1$ the surface is of the form shown in figure 10. It
extends to infinity and has an discontinuity at $x=0$. If $A\varkappa =1$ it
becomes the cuspidal edge ($\frac{dZ}{dr}=0$ at $x=0$) and the corresponding
surface, which can be seen in figure 11, consists of two parts ( $x<0$ and $%
x>0$) where both are just the pseudosphere discussed in section 3. In the
case of $A\varkappa >1$ the surface of revolution is built from two
disconnected intervals of the parameter $x$, namely $-\infty <x<-\frac
1\varkappa \ln \left( A\varkappa \right) $ and $\frac 1\varkappa \ln \left(
A\varkappa \right) <x<\infty $. They have cuspidal edges at $x=\pm \frac
1\varkappa \ln \left( A\varkappa \right) $. Each of these parts is nothing
but the pseudosphere. For the solutions $\Psi \left( x\right) $ with $E>0$
the corresponding surfaces look similar to the case of free motion with $E>0$%
.

In a similar manner one can construct surfaces associated with a potential
which is the sum of several $\delta $-functions. The presence of $\delta $%
-functions allows us to describe surfaces with constant Gaussian curvature
except at some points which correspond to edges. The Dirac comb potential

\begin{equation}
u\left( x\right) =\alpha \sum_{n=-\infty }^\infty \delta \left( x-na\right)
\label{e5.4}
\end{equation}

is of particular interest. In solid state physics it is used as a model
potential with periodic structure which leads to energy bands for the
stationary states. In our approach it allows to describe surfaces with an
infinite number of cuspidal edges. If $\alpha >0$ the Dirac comb generates
for $E>0$ surfaces which look like pieces of the form as in figure 3 glued
together. Between the points $x=na$ they have constant positive Gaussian
curvature while at these points the curvature is negative and infinite.

The case $\alpha <0$ and $E<0$ corresponds to the surface which consists of
an infinite sequence of the parts presented in the figures 4 and 5. So we
have the ''pseudospherical' surface with an infinite set of edges. Note,
that if we exclude the points $x=na$ then the rest is the disconnected
pseudospherical surface.

\section{Surfaces Associated with the Harmonic Oscillator}

The following example is the famous harmonic oscillator potential, namely 
\begin{equation}
u\left( x\right) =x^2.  \label{e6.1a}
\end{equation}

It has an infinite set of stationary states with the energies

\begin{equation}
E_n=2\left( n+\frac 12\right) \;\;,\;\;n=0,1,2,...  \label{e6.1b}
\end{equation}

and the wavefunctions

\begin{equation}
\Psi _n\left( x\right) =A_n\exp \left( -\frac{x^2}2\right) H_n\left(
x\right) ,  \label{e6.1}
\end{equation}

where $H_n$ are the Hermite polynomials and $A_n$ are arbitrary real
constants \cite{23}. For the ground state ($n=0$), the first ($n=1$) and the
second ($n=2$) excited state the corresponding wavefunctions take the form

\begin{eqnarray}
\Psi _0\left( x\right) &=&A_0\exp \left( -\frac{x^2}2\right)  \label{e6.1c}
\\
&&  \nonumber \\
\Psi _1\left( x\right) &=&A_1x\exp \left( -\frac{x^2}2\right)  \label{e6.1d}
\\
&&  \nonumber \\
\Psi _2\left( x\right) &=&A_2[4x^2-2]\exp \left( -\frac{x^2}2\right) ,
\label{e6.1e}
\end{eqnarray}

respectively. In the figures 12, 13 and 14 one can see the induced surfaces
for a choice of the constants $A_n$ which satisfies the constraint $\left|
\Psi _x\right| <1$ for all $x$. Figure 15 shows a surface connected to the
ground state where $A_0$ was chosen in such a way that at the points $x=\pm
1 $ the tangent plane on the surface is parallel to the $X$-$Y$-plane, i.e. $%
\left| \Psi _x\right| =1$.

\section{Effective One-Dimensional Coulomb Potential}

With the help of the substitution

\begin{equation}
R\left( r\right) =\frac{\Psi \left( r\right) }r  \label{e7.1a}
\end{equation}

the radial Schr\"{o}dinger equation for the hydrogen atom can be reduced to
the one-dimensional equation ( \cite{23}, \S 36)

\begin{equation}
-\Psi _{xx}+\left[ \frac{l\left( l+1\right) }{x^2}-\frac 2x\right] \Psi
=2E\Psi ,  \label{e7.1}
\end{equation}

where $l$ is the angular momentum and we substituted $r$ by $x$. The
functions $\Psi \left( x\right) $ corresponding to the bound states with
energies

\begin{equation}
E_n=-\frac 1{2n^2}  \label{e7.1b}
\end{equation}

are given in the form

\begin{equation}
\Psi _{n,l}\left( x\right) =A_nx^{l+1}\exp (-\frac xn)L_{n+l}^{2l+1}\left( 
\frac{2r}n\right) .  \label{e7.2}
\end{equation}

In (\ref{e7.2}) the functions $L_{n+l}^{2l+1}$are the Laguerre polynomials
and $A_n$ are as usual arbitrary real constants. As can be seen from (\ref
{e7.1}) the Gaussian curvature now has pole type singularities. For the
ground state ($n=1$, $l=0$) we thus have

\begin{eqnarray}
K_{1,0}\left( x\right) &=&\frac 2x-\frac 12  \label{e7.2a} \\
&&  \nonumber \\
\Psi _{1,0}\left( x\right) &=&A_1x\exp \left( -x\right) .  \label{e7.2b}
\end{eqnarray}

Figure 16 gives an impression of the surface. For $n=2$ we have the bound
states with $l=0$ ,

\begin{eqnarray}
K_{2,0}\left( x\right) &=&\frac 2x-\frac 18  \label{e7.2c} \\
&&  \nonumber \\
\Psi _{2,0}(x) &=&A_2x\left( 1-\frac x2\right) \exp \left( -\frac x2\right) ,
\label{e7.2d}
\end{eqnarray}

and $l=1$ corresponding to

\begin{eqnarray}
K_{2,1}\left( x\right) &=&\frac 2x-\frac 2{x^2}-\frac 18  \label{e7.3} \\
&&  \nonumber \\
\Psi _{2,1}\left( x\right) &=&A_2x^2\exp \left( -\frac x2\right) .
\label{e7.4}
\end{eqnarray}

The figures 17 and 18 show the surfaces of revolution, respectively.

\section{Soliton Surfaces via Bargmann Potentials}

A wide class of solvable potentials for the Schr\"{o}dinger equation is
given by the multi-Bargmann potentials (see e.g. \cite{8} - \cite{10}). The
corresponding surfaces of revolution can be represented as

\begin{eqnarray}
K\left( x\right) &=&\lambda ^2+2(\ln \det D)_{xx}  \label{e8.1} \\
&&  \nonumber \\
H\left( x\right) &=&Re\left[ A\exp \left( -i\lambda x\right) \left\{
1+\sum_{k=1}^N\frac{\det D^{\left( k\right) }}{\det D}\frac{\exp \left(
-\lambda _kx\right) }{\lambda _k+i\lambda }\right\} \right] ,  \label{e8.2}
\end{eqnarray}

where $E=\lambda ^2$ and the matrix elements of the $N\times N$ matrix $D$
read

\begin{equation}
D_{kl}=\delta _{kl}+\frac{\mu _k\exp \left( -\left[ \lambda _k+\lambda
_l\right] x\right) }{\lambda _k+\lambda _l}  \label{e8.3}
\end{equation}

for $k,l=1,...,N$. $N$ is an arbitrary integer, $\lambda _k$ and $\mu _k$
are arbitrary real constants and the matrix elements of the matrix $%
D^{\left( k\right) }$ are given by (\ref{e8.3}) where the last column is
substituted by the column $-\beta _k\exp \left( -\lambda _kx\right) $ ($%
k=1,...,N$). The $N$-Bargmann potential has $N$ bound states with the
energies $-\lambda _1^2$, $-\lambda _2^2$, ..., $-\lambda _N^2$ ($\lambda
_1<\lambda _2<...<\lambda _N$). $A$ is an arbitrary amplitude.

Let us first consider the simplest Bargmann potential, i.e. $N=1$. Depending
on the choice of the parameter $\lambda $ we get three different types of
surfaces. We will only treat the case when $A$ is chosen in such a way that
the condition $\left| H_x\right| <1$ is satisfied which corresponds to
regular surfaces of revolution.

\begin{itemize}
\item[1.]  For the bound state we put $\lambda =i\lambda _1$ which leads to 
\begin{eqnarray}
K\left( x\right) =-\lambda _1^2+\frac{2\lambda _1^2}{\cosh ^2\left( \lambda
_1x\right) }  \label{e8.4} \\
\nonumber \\
H\left( x\right) =\frac A{\cosh \left( \lambda _1x\right) },  \label{e8.5}
\end{eqnarray}

where $A$ is once again an arbitrary real constant. The corresponding
surface looks like a bubble which decays exponentially fast into strings
tending to infinity. Figure 19 shows the surface.

\item[2.]  The case of zero energy, i.e. $\lambda =0$ leads to the
corresponding surface of revolution plotted in figure 20. Gaussian curvature
and metric take the form 
\begin{eqnarray}
K\left( x\right) =\frac{2\lambda _1^2}{\cosh ^2\left( \lambda _1x\right) }
\label{e8.6} \\
\nonumber \\
H\left( x\right) =A\tanh \left( \lambda _1x\right) .  \label{e8.7}
\end{eqnarray}

\item[3.]  If we want to consider a positive energy we have to choose $%
\lambda $ real. The associated surface of revolution then represents a
scattering state which, up to a slight decrease of the amplitude in the area
where the potential has its maximum, is similar to the free motion case. In
that case we have 
\begin{eqnarray}
K\left( x\right) =\lambda ^2+\frac{2\lambda _1^2}{\cosh ^2\left( \lambda
_1x\right) }  \label{e8.8} \\
\nonumber \\
H\left( x\right) =A[\lambda \sin \left( \lambda x\right) +\lambda _1\cos
\left( \lambda x\right) \tanh \left( \lambda _1x\right) ].  \label{e8.9}
\end{eqnarray}

Figure 21 shows the surface for a special choice of parameters. Note, that
far from the potential well the surface has the same periodic form on both
sides. This corresponds exactly to the transparency of the Bargmann
potential.
\end{itemize}

In the case $N=2$ we end up with a potential having two wells separated by a
distance which can be adjusted freely simply by fixing the corresponding
parameters. We will omit the formulas for Gaussian curvature and the metric
here, they can be derived straightforward from (\ref{e8.1}) and (\ref{e8.2}%
), for details we refer to section 10. Two figures might be sufficient to
illustrate the situation, namely in figure 22 we see a surface corresponding
to the ground state of the 2-soliton potential, whereas figure 23 shows the
excited bound state surface.

\section{Deformations of Surfaces Preserving the Gaussian Curvature}

In this section we will consider deformations of the surfaces of revolution
constructed above. The simplest type of deformations is given by rescaling
the metric, i.e.

\begin{equation}
H\left( x\right) \rightarrow H^{\prime }\left( x\right) =g\left( t\right)
H\left( x\right) ,  \label{e9.1}
\end{equation}

where $g\left( t\right) $ is an arbitrary function. The Gaussian curvature
is invariant under such deformations, but the shape of the surfaces in $R^3$
may change drastically.

Let us start with the surfaces associated with the free motion (section 3)
at $E>0$ in the case $Ak<1$. Let $H_x\left( t=0\right) <1$ and $g\left(
t\right) $ be a monotonically increasing function with $g\left( t=0\right)
=1 $. Hence at $t=0$ we have the surfaces of the type depicted in figure 2.
For $t>0$ the height of each segment increases and at the moment $t_0$
defined by $Akg\left( t_0\right) =1$ the surface becomes the sphere. For $%
t>t_0$ there will be cuspidal edges and the surface changes its shape to
that given in figure 3. For monotonically decreasing functions $g\left(
t\right) $ the deformations of the surface proceed in the opposite
direction. Evidently, for deformations given by functions such that $\left|
g\left( t\right) \right| <1$ for all $t$ the shape of the surface is changed
unessentially.

For the pseudospherical surfaces the deformations (\ref{e9.1}) look simpler.
For the surfaces of figure 4 and 5 this type of deformations changes, in
essence, only the distance between the edges. For the pseudosphere of figure
6 the deformation (\ref{e9.1}) changes just the position of the edge which
can always be shifted to the origin by a simple redefinition of $x$. So, the
pseudosphere is, in fact, invariant with respect to deformations (\ref{e9.1}%
) for any smooth function $g\left( t\right) $.

The behavior of other surfaces of revolution considered in the previous
sections under the deformations of the type (\ref{e9.1}) is similar. For
regular surfaces ($|H_x|<1$) corresponding to bound states (\ref{e9.1}) will
disconnect the surface. These cuts are developing, evidently, at those
points where $|H_x|$ has maximums. For instance, for the surface generated
by the bound state of the 1-soliton potential with $H=A/\cosh \left( \lambda
_1x\right) $ the function $H_x$ has extrema at the points $x_{1,2}=\pm \frac
1{\lambda _1}{\rm arccosh}\left( \sqrt{2}\right) $. Thus, for a
monotonically increasing function $g\left( t\right) $ as $t\rightarrow t_0$,
where $t_0$ is the time when $|H_x|$ becomes equal to one, the profile of
the surface near the points $x_{1,2}=\pm 1$ becomes parallel to the $X$-$Y$%
-plane at $t_0$ and later the surfaces gets disconnected Figure 24 shows the
behavior for a special choice of the function $g\left( t\right) $. As $%
t\rightarrow \infty $ the outer ring becomes thinner and finally disappears
whereupon the whole surface is converted into the pseudosphere. Note, that
for deformations (\ref{e9.1}) the splitting of surfaces after some finite
time $\Delta t$ when starting from regular ones is a general feature. The
lap $\Delta t$, of course, is determined by the function $g\left( t\right) $
and the smallest difference $|H_x|-1$.

A quite obvious generalization of the deformations (\ref{e9.1}) is of the
form

\begin{equation}
\Psi (x)\longrightarrow \Psi ^{\prime }(x,t)=A(t)\Psi (x)+B(t)\widetilde{%
\Psi }(x),  \label{e9.2}
\end{equation}

where $\widetilde{\Psi }(x)$ is the linearly independent solution of the
Schr\"{o}dinger equation with respect to $\Psi (x)$ and $A(t)$ and $B(t)$
are two arbitrary functions which satisfy $A\left( 0\right) =1$ and $B\left(
0\right) =0$. In the general case $\widetilde{\Psi }(x)$ is related to $\Psi
(x)$ according to

\begin{equation}
\widetilde{\Psi }(x)=\Psi (x)\int^x\frac{dx^{\prime }}{\Psi ^2(x^{\prime })}
\label{e9.2a}
\end{equation}

whereupon one can rewrite (\ref{e9.2}) as

\begin{equation}
\Psi (x)\longrightarrow \Psi ^{\prime }(x,t)=A(t)\Psi (x)+B(t)\Psi (x)\int^x%
\frac{dx^{\prime }}{\Psi ^2(x^{\prime })}.  \label{e9.3}
\end{equation}

The deformations (\ref{e9.3}) also do not change the Gaussian curvature but
their action on the profile of the surfaces is stronger than that of the
transformations (\ref{e9.1}).

For the standard sphere (figure 1) the linearly independent solution is $%
\widetilde{\Psi }(x)=\frac 1k\sin \left( kx\right) $ so (\ref{e9.3}) reads
explicitly

\begin{equation}
\Psi ^{\prime }(x,t)=\frac{A(t)}k\cos (kx)+\frac{B(t)}k\sin \left( kx\right)
\label{e9.4}
\end{equation}

or

\begin{equation}
\Psi ^{\prime }(x,t)=\frac{\sqrt{A^2\left( t\right) +B^2\left( t\right) }}%
k\cos [kx-\varphi \left( t\right) ]  \label{e9.5}
\end{equation}

where

\begin{equation}
\varphi \left( t\right) =\arccos \frac{A\left( t\right) }{\sqrt{A^2\left(
t\right) +B^2\left( t\right) }}.  \label{e9.5a}
\end{equation}

Thus the deformation of the sphere is given by a motion along $x$ with the
velocity $\varphi \left( t\right) $ and a change of shape to that type shown
in figure 2 as long as $\left| \Psi _x(x,t)\right| <1$ for all $t$. The
situation is similar for other spherical surfaces.

For the pseudospherical surfaces of figure 4 and 5 the transformation (\ref
{e9.3}) is of the form

\begin{equation}
\Psi ^{\prime }(x,t)=A\left( t\right) \cosh (\lambda x)+B(t)\sinh \left(
\lambda x\right) .  \label{e9.6}
\end{equation}

Basically, this results again in shifting and rescaling the surface. The
only significant change of the shape concerns the distance between the two
edges.

For the pseudosphere the situation is quite different. In this case the
linear independent solution of the Schr\"{o}dinger equation is $\widetilde{%
\Psi }(x)=\frac 1\lambda \exp \left( \lambda x\right) $ leading to a
deformation of the form

\begin{equation}
\Psi ^{\prime }(x,t)=\frac{A\left( t\right) }\lambda \exp \left( -\lambda
x\right) +\frac{B\left( t\right) }\lambda \exp \left( \lambda x\right) .
\label{e9.7}
\end{equation}

Since now $H_x$ is growing not only at $x<0$ but also for $x>0$ there are
for any given $B(t)$ at any time $t\neq 0$ two cuspidal edges on the
deformed surface. So the deformation (\ref{e9.7}) creates immediately after
the initial moment $t=0$ an additional edge on the pseudosphere which
thereby takes a shape similar to those of figures 4 and 5 depending on the
explicit form of the functions $A\left( t\right) $ and $B\left( t\right) $.

A similar behavior can be observed for surfaces associated with bound states
except the infinite well potential. For bound states the wavefunctions are
decreasing at $\left| x\right| \rightarrow \infty $. This, however, implies
that the linearly independent solution is increasing at infinity. Hence the
function $\Psi (x,t)$ defined by (\ref{e9.3}) always has an additional edge
for any $t\neq 0$ for any function $B\left( t\right) $. Therefore, regular
surfaces connected to bound states are unstable with respect to deformations
of the form (\ref{e9.3}).

A very particular class of deformations (\ref{e9.3}) can be obtained if we
choose the following form for the functions $A\left( t\right) $ and $B\left(
t\right) $

\begin{equation}
A=\frac 1{\sqrt{-2T_t(t)}}\;,\;B=\frac{T(t)}{\sqrt{-2T_t(t)}}  \label{e9.8}
\end{equation}

where $T(t)$ is an arbitrary function and $T_t(t)$ is the derivative of $%
T(t) $ with respect to $t$. This choice leads to an evolution for the
Schr\"{o}dinger equation (\ref{e2.10}) recently discussed in \cite{21} which
does not change the potential. In terms of the variable

\begin{equation}
q=-2\ln \Psi  \label{e9.8a}
\end{equation}

such an evolution in time is governed by the Liouville equation \cite{21}

\begin{equation}
q_{xt}=e^q.  \label{e9.9}
\end{equation}

The well-known general solution of the Liouville equation

\begin{equation}
q\left( x,t\right) =\ln \frac{2T_t\left( t\right) S_x\left( x\right) }{%
[T\left( t\right) +S\left( x\right) ]^2}  \label{e9.10}
\end{equation}

implies

\begin{equation}
\Psi \left( x,t\right) =\frac{T\left( t\right) +S\left( x\right) }{\sqrt{%
2T_t\left( t\right) S_x\left( x\right) }}  \label{e9.11}
\end{equation}

where $T\left( t\right) $ and $S\left( x\right) $ are arbitrary functions.
Demanding at $t=0$ that $T\left( 0\right) =0$ and $T_t(0)=-1/2$ one has

\begin{equation}
\Psi \left( x,0\right) =\frac{S\left( x\right) }{\sqrt{-S_x\left( x\right) }}
\label{e9.11a}
\end{equation}

which after integration leads to

\begin{equation}
S\left( x\right) =\left( \int^x\frac{dx^{\prime }}{\Psi ^2\left( x^{\prime
},0\right) }\right) ^{-1}.  \label{e9.11b}
\end{equation}

Substituting this expression into (\ref{e9.11}) one gets

\begin{equation}
\Psi \left( x,t\right) =\frac{\Psi \left( x,0\right) }{\sqrt{-2T_t\left(
t\right) }}+\frac{T\left( t\right) }{\sqrt{-2T_t\left( t\right) }}\Psi
\left( x,0\right) \int^x\frac{dx^{\prime }}{\Psi ^2\left( x^{\prime
},0\right) },  \label{e9.12}
\end{equation}

i.e. nothing but (\ref{e9.3}) with $A$ and $B$ given by (\ref{e9.8}). Thus
the Liouville type deformations have properties similar to those discussed
above. Figure 25 illustrates the situation for the ground state of the
1-soliton potential.

\section{Deformations via the KdV Equation}

A completely different class of deformations of surfaces of revolution is
given by the KdV equation which governs the isospectral deformations of the
Schr\"{o}dinger equation (\ref{e2.10}). It reads (see e.g. \cite{8} - \cite
{10})

\begin{equation}
u_t+u_{xxx}-6uu_x=0  \label{e10.1}
\end{equation}

while the wavefunction $\Psi $ evolves according to the linear equation

\begin{equation}
\Psi _t+4\Psi _{xxx}-6u\Psi _x-3u_x\Psi =0.  \label{e10.2}
\end{equation}

Eliminating $u$ from (\ref{e10.2}) with the use of (\ref{e2.10}) one gets
the equation for $\Psi $ only, namely

\begin{equation}
\Psi _t+\Psi _{xxx}-6E\Psi _x-3\frac{\Psi _x\Psi _{xx}}\Psi =0.
\label{e10.3}
\end{equation}

The KdV equation is integrable by the IST method. Using this method one can
analyze this equation in detail (see e.g. \cite{8} - \cite{10}). In
particular, one reduces the solution of the nonlinear initial value problem $%
u\left( x,0\right) \rightarrow u\left( x,t\right) $ to a set of linear
problems. For a wide class of initial data it is possible to calculate exact
asymptotics at $t\rightarrow \infty $. The IST method provides us with
explicit formulas for the multi-soliton solutions. The KdV equation has a
number of remarkable properties : infinite symmetry group, Darboux and
B\"{a}cklund transformations and an infinite set of integrals of motion
given by

\begin{equation}
Q_n=2\pi \int_{-\infty }^\infty dxC_{2n+1}(x),  \label{e10.4}
\end{equation}

where the densities $C_n$ are calculated via the recurrent relations

\begin{eqnarray}
C_{n+1} &=&C_{n_x}+\sum_{k=1}^{n-1}C_kC_{m-k}\;\;,\;\;n=1,2,...
\label{e10.5a} \\
&&  \nonumber \\
C_1 &=&-u\left( x,t\right) .  \label{e10.5}
\end{eqnarray}

All $C_{2n}$ are total derivatives. So the nontrivial integrals are given by
(\ref{e10.4}).

The KdV deformations of surfaces of revolution are governed by the equations
(\ref{e10.1}) - (\ref{e10.3}) with the substitution $K=E-u$, i.e. the
Gaussian curvature evolves according to the equation

\begin{equation}
K_t-6EK_x+K_{xxx}+6KK_x=0  \label{e10.7}
\end{equation}

and the metric depends on time via

\begin{equation}
H_t-6EH_x+4H_{xxx}+6KH_x+3K_xH=0.  \label{e10.8}
\end{equation}

The KdV deformations of surfaces inherit all remarkable properties of the
KdV equation. First, one is able to linearize the initial value problem $%
\left\{ K\left( x,0\right) ,H\left( x,0\right) \right\} \rightarrow \left\{
K\left( x,t\right) ,H\left( x,t\right) \right\} $ for the deformations of
the surfaces. Moreover, it is possible to find exact asymptotic expressions
for the Gaussian curvature and the metric as $t\rightarrow \infty $ and,
finally, an infinite set of deformations of surfaces is given in terms of
explicit formulas, namely

\begin{eqnarray}
K\left( x\right) &=&\lambda ^2+2(\ln \det D)_{xx}  \label{e10.9a} \\
&&  \nonumber \\
H\left( x\right) &=&Re\left[ A\exp \left( -i\lambda x\right) \left\{
1+\sum_{k=1}^N\frac{\det D^{\left( k\right) }}{\det D}\frac{\exp \left(
-\lambda _kx\right) }{\lambda _k+i\lambda }\right\} \right] ,  \label{e10.9}
\end{eqnarray}

where $E=\lambda ^2$ and the matrix elements of the $N\times N$ matrix $D$
read

\begin{equation}
D_{kl}=\delta _{kl}+\frac{\exp \left( -\left[ \lambda _k+\lambda _l\right]
x+8\lambda _k^3t+\gamma _k\right) }{\lambda _k+\lambda _l}.  \label{e10.10}
\end{equation}

The matrix elements of the matrices $D^{\left( k\right) }$ are given by (\ref
{e10.10}) with the substitution of the last column by the column $-\exp
\left( -\lambda _kx+8\lambda _k^3t+\gamma _k\right) $ for $k=1,...,N.$ The
parameters $\lambda _k$ are real constants and $\gamma _k$ are arbitrary
phases. $A$ is an arbitrary amplitude. According to section 8 where we
discussed soliton surfaces we can now investigate their evolution under KdV
deformation. To this end, we will restrict ourselves to three examples of
regular surfaces, i.e. to an appropriate choice of $A$.

\begin{itemize}
\item[1)]  The surface generated by the bound state of the 1-soliton
potential evolves in time according to the following Gaussian curvature and
metric 
\begin{eqnarray}
K\left( x\right) =-\lambda _1^2+\frac{2\lambda _1^2}{\cosh ^2\left( \lambda
_1x-4\lambda _1^3t+\gamma _1\right) }  \label{e10.11} \\
\nonumber \\
H\left( x\right) =\frac A{\cosh \left( \lambda _1x-4\lambda _1^3t+\gamma
_1\right) }.  \label{e10.12}
\end{eqnarray}

The deformation (\ref{e10.11}) and (\ref{e10.12}) is nothing but the uniform
motion of the 1-soliton surface along the $Z$-axis with the constant
velocity $4\lambda _1^3$ without a change of motion (cf. figure 19).

\item[2)]  In the case of two solitons we can write the potential in the
form ($\lambda _1<\lambda _2$)

\begin{equation}
u\left( x,t\right) =\frac{W_1}{W_2}  \label{e10.12a}
\end{equation}

with

\begin{eqnarray}
W_1 &=&2(\lambda _2^2-\lambda _1^2)[\lambda _2^2\cosh ^2\xi _1\left( \sinh
^2\xi _2-\cosh ^2\xi _2\right)  \nonumber \\
&&\;\;\;\;\;\;\;\;\;\;\;\;\;\;+\lambda _1^2\sinh ^2\xi _2\left( \sinh ^2\xi
_1-\cosh ^2\xi _1\right) ]  \label{e10.12b} \\
&&  \nonumber \\
W_2 &=&\left[ \lambda _2\cosh \xi _1\cosh \xi _2-\lambda _1\sinh \xi _1\sinh
\xi _2\right] ^2  \label{e10.12c}
\end{eqnarray}

and

\begin{equation}
\xi _{1,2}=\lambda _{1,2}x-4\lambda _{1,2}^3t+\gamma _{1,2}.  \label{e10.12d}
\end{equation}

Thus for the ground state the Gaussian curvature and the metric take the
form 
\begin{eqnarray}
K\left( x\right) &=&-\lambda _2^2-u\left( x,t\right) \;  \label{e10.12e} \\
&&  \nonumber \\
H\left( x\right) &=&A\frac{\lambda _2(\lambda _2^2-\lambda _1^2)\cosh \xi _1%
}{\lambda _2\cosh \xi _1\cosh \xi _2-\lambda _1\sinh \xi _1\sinh \xi _2}.
\label{e10.12f}
\end{eqnarray}

while the surface corresponding to first excited state is described by 
\begin{eqnarray}
K\left( x\right) =-\lambda _1^2-u\left( x,t\right) \;  \label{e10.12g} \\
\nonumber \\
H\left( x\right) =A\frac{\lambda _1(\lambda _1^2-\lambda _2^2)\sin \xi _2}{%
\lambda _2\cosh \xi _1\cosh \xi _2-\lambda _1\sinh \xi _1\sinh \xi _2}.
\label{e10.12h}
\end{eqnarray}

For the ground state we see one bubble moving along the $Z$-axis which in
contrast to the 1-soliton case slightly changes its shape. The first excited
state, however, mimics the 2-soliton interaction of the KdV equation.
Figures 26 and 27 illustrate the deformation of the two surfaces of
revolution.
\end{itemize}

For the KdV deformations the Gaussian curvature evolves in time but there is
an infinite set of integral characteristic of the surfaces which are
preserved. They are of the form (\ref{e10.4}) with the substitution $%
u\rightarrow E-K$. For $N$-soliton deformations these integral
characteristics are (see e.g. \cite{8})

\begin{equation}
Q_n=2\pi \frac{2^{2n}}{2n-1}\sum_{k=1}^N\lambda
_k^{2n-1}\;\;,\;\;n=1,2,3,...\;.  \label{e10.13}
\end{equation}

It is known that any initial data which obey the constraint (see \cite{8}-%
\cite{10})

\begin{equation}
\int_{-\infty }^\infty dx\left( 1+\left| x\right| \right) \left| u\left(
x\right) \right| <\infty  \label{e10.14}
\end{equation}

decompose asymptotically ($t\rightarrow \infty $) under the KdV evolution
into pure solitons. In terms of surfaces this means that asymptotically we
end up with soliton surfaces. To obtain a regular surface for all times the
amplitude $A$ has to be fixed in an appropriate way, but since during the
evolution the wavefunction is bounded it is always possible to adjust $A$
such that $\left| H_x\right| <1$ for all $t$. Moreover, the KdV deformations
tend to smooth out the initial surfaces. Indeed, let us take the $\delta $%
-function as the initial data, i.e. $u\left( x,0\right) =-2\varkappa \delta
\left( x\right) $. The corresponding pseudosphere has an edge at $x=0$.
Under the KdV evolution this initial data converts asymptotically into one
soliton (see e.g. \cite{10}). Thus, the KdV deformation converts the
pseudosphere figure 11 into the soliton bubble, figure 19, as $t\rightarrow
\infty $. In contrast, starting from the regular 1-soliton surface the
deformations discussed in section 9, namely (\ref{e9.1}) and (\ref{e9.3})
create at finite time or immediately singularities, i.e. edges. So, there is
a crucial difference between the KdV deformations and those which preserve
the Gaussian curvature. The former tend to smooth out the surfaces while the
latter create singularities.

\section{Acknowledgment}

One of us (RB) would like to thank the Department of Physics of the
University of Lecce for kind hospitality. This work was supported by the
European Community in terms of the research program {\em Human Capital and
Mobility}.

\newpage

\section*{Figure Captions}

\begin{tabular}{p{2cm}p{13cm}}
Figure 1: \hfill & Free motion, $\Psi \left( x\right) =A\cos (kx)$; $A=2$, $%
k=0.5$ \hfill \\ 
\hfill & \hfill \\ 
Figure 2: \hfill & Free motion, $\Psi \left( x\right) =A\cos (kx)$; $A=1.8$, 
$k=0.5$ \hfill \\ 
\hfill & \hfill \\ 
Figure 3: \hfill & Free motion, $\Psi \left( x\right) =A\cos (kx)$; $A=2.2$, 
$k=0.5$ \hfill \\ 
\hfill & \hfill \\ 
Figure 4: \hfill & Free motion, $\Psi \left( x\right) =A\cosh (\lambda x)$; $%
A=0.5$, $\lambda =1$ \hfill \\ 
\hfill & \hfill \\ 
Figure 5: \hfill & Free motion, $\Psi \left( x\right) =A\sinh (\lambda x)$; $%
A=0.5$, $\lambda =1$ \hfill \\ 
\hfill & \hfill \\ 
Figure 6: \hfill & Free motion, $\Psi \left( x\right) =A\exp (-\lambda x)$; $%
A=1$, $\lambda =1$ \hfill \\ 
\hfill & \hfill \\ 
Figure 7: \hfill & Infinite well potential, $\Psi \left( x\right) =A\sin
\left(\frac{nx}a\right) $; $a=1$, $n=1,..,4$, $A=\frac an$ \hfill \\ 
\hfill & \hfill \\ 
Figure 8: \hfill & Infinite well potential, $\Psi \left( x\right) =A\sin
\left(\frac{nx}a\right) $; $a=1$, $n=1,..,4$, $A=\frac{5a}{6n}$ \hfill \\ 
\hfill & \hfill \\ 
Figure 9: \hfill & Infinite well potential, $\Psi \left( x\right) =A\sin
\left(\frac{nx}a\right) $; $a=1$, $n=1,..,4$, $A=\frac{6a}{5n}$ \hfill \\ 
\hfill & \hfill \\ 
Figure 10: \hfill & $\delta $-function potential, $\Psi \left( x\right) =
A\exp\left( -k\left| x\right| \right) $; $A=0.5$, $k=1$ \hfill \\ 
\hfill & \hfill \\ 
Figure 11: \hfill & $\delta $-function potential, $\Psi \left( x\right) =
A\exp\left( -k\left| x\right| \right) $; $A=1$, $k=1$ \hfill \\ 
\hfill & \hfill \\ 
Figure 12: \hfill & Harmonic oscillator potential, $\Psi \left( x\right) =
A\exp\left( -\frac{x^2}2\right) $; $A=1$ \hfill \\ 
\hfill & \hfill \\ 
Figure 13: \hfill & Harmonic oscillator potential, $\Psi \left( x\right) =
Ax\exp\left( -\frac{x^2}2\right) $; $A=0.8$ \hfill \\ 
\hfill & \hfill \\ 
Figure 14: \hfill & Harmonic oscillator potential, $\Psi \left( x\right)
=A(4x^2-2)\exp \left( -\frac{x^2}2\right) $; $A=0.2$ \hfill \\ 
\hfill & \hfill \\ 
Figure 15: \hfill & Harmonic oscillator potential, $\Psi \left( x\right) =
A\exp\left( -\frac{x^2}2\right) $; $A=e^{1/2}$ \hfill \\ 
\hfill & \hfill \\ 
Figure 16: \hfill & Effective Coulomb potential, $\Psi \left( x\right) =
Ax\exp\left(-x\right) $; $A=0.9$ \hfill \\ 
\hfill & \hfill \\ 
Figure 17: \hfill & Effective Coulomb potential, $\Psi \left( x\right) =
Ax\left(1-\frac{x}{2}\right) \exp\left(-\frac{x}{2}\right) $; $A=0.5$ \hfill
\end{tabular}
\newpage 
\begin{tabular}{p{2cm}p{13cm}}
Figure 18: \hfill & Effective Coulomb potential, $\Psi \left( x\right) =
Ax^2\exp\left(-\frac{x}{2}\right) $; $A=0.16$ \hfill \\ 
Figure 19: \hfill & 1-soliton potential, ground state, \hfill \\ 
\hfill & \hfill \\ 
\hfill & $\Psi \left(x\right) = \frac{\displaystyle A} {\displaystyle%
\cosh\left(\lambda_1 x\right)}$; $A=1$, $\lambda_1=1$ \hfill \\ 
\hfill & \hfill \\ 
Figure 20: \hfill & 1-soliton potential, zero energy, \hfill \\ 
\hfill & \hfill \\ 
\hfill & $\Psi \left(x\right) = A\tanh\left(\lambda_1 x\right)$; $A=0.5$, $%
\lambda_1=1$ \hfill \\ 
\hfill & \hfill \\ 
Figure 21: \hfill & 1-soliton potential, positive energy, \hfill \\ 
\hfill & \hfill \\ 
\hfill & $\Psi \left(x\right) = A\left[\lambda\sin\left(\lambda x\right) +
\lambda_1\cos\left(\lambda x\right)\tanh\left(\lambda_1 x\right) \right]$; $%
A=0.4$, $\lambda_1=1$, $\lambda=1$ \hfill \\ 
\hfill & \hfill \\ 
Figure 22: \hfill & 2-soliton potential, ground state, \hfill \\ 
\hfill & \hfill \\ 
\hfill & $\Psi\left(x\right)=A \frac{\displaystyle\lambda_2\left(%
\lambda_2^2-\lambda_1^2\right)\cosh\xi_1} {\displaystyle\lambda_2\cosh\xi_1%
\cosh\xi_2-\lambda_1\sinh\xi_1\sinh\xi_2}$, \hfill \\ 
\hfill & \hfill \\ 
\hfill & $\xi_k=\lambda_k x -4\lambda^2t+\gamma_k$; $A=1$, $\lambda_1=0.5$, $%
\lambda_2=1$, $\gamma_1=1$, $\gamma_2=-1$, $t=-2$ \hfill \\ 
\hfill & \hfill \\ 
Figure 23: \hfill & 2-soliton potential, excited state, \hfill \\ 
\hfill & \hfill \\ 
\hfill & $\Psi\left(x\right)=A \frac{\displaystyle\lambda_1\left(%
\lambda_1^2-\lambda_2^2\right)\sinh\xi_2} {\displaystyle\lambda_2\cosh\xi_1%
\cosh\xi_2-\lambda_1\sinh\xi_1\sinh\xi_2}$, \hfill \\ 
\hfill & \hfill \\ 
\hfill & $\xi_k=\lambda_k x -4\lambda^2t+\gamma_k$; $A=1$, $\lambda_1=0.5$, $%
\lambda_2=1$, $\gamma_1=1$, $\gamma_2=-1$, $t=-2$ \hfill \\ 
\hfill & \hfill \\ 
Figure 24: \hfill & 1-soliton potential, deformation (\ref{e9.1}) with $%
g(t)=1+t$; \hfill \\ 
\hfill & $\lambda_1 = 1$, $t=0,1,1.1,1.5$ \hfill \\ 
\hfill & \hfill \\ 
Figure 25: \hfill & 1-soliton potential, deformation (\ref{e9.12}) with $%
T(t)=-\frac{t}{2}$; \hfill \\ 
\hfill & $\lambda_1=1$, $t=0,0.1,0.5,1$ \hfill \\ 
\hfill & \hfill \\ 
Figure 26: \hfill & 2-soliton potential, ground state, KdV deformation;
\hfill \\ 
\hfill & $t=-1.5,-1,-0.5,0$ \hfill
\end{tabular}
\newpage 
\begin{tabular}{p{2cm}p{13cm}}
Figure 27: \hfill & 2-soliton potential, excited state, KdV deformation;
\hfill \\ 
\hfill & $t=-1.5,-1,-0.5,0$ \hfill
\end{tabular}

\end{document}